\documentclass[prb,twocolumn,showpacs]{revtex4}
\usepackage{amssymb}
\usepackage{graphics}
\usepackage{graphicx}
\usepackage{bm}

\renewcommand{\ss}{\scriptscriptstyle}
\input{tcilatex}

\begin{document}

\title{Temporal stimulated intersubband emission of photoexcited electrons }
\author{F.T. Vasko}
\email{ftvasko@yahoo.com}
\affiliation{Institute of Semiconductor Physics, NAS Ukraine, Pr. Nauki 41, Kiev, 03028,
Ukraine}
\author{A. Hern{\'{a}}ndez-Cabrera}
\email{ajhernan@ull.es}
\author{P. Aceituno}
\affiliation{Dpto. F{\'\i}sica B\'{a}sica, Universidad de La Laguna, La Laguna,
38206-Tenerife, Spain}
\date{\today }

\begin{abstract}
We have studied the transient evolution of electrons distributed over two
levels in a wide quantum well, with the two levels below the optical phonon
energy, after an ultrafast interband excitation and cascade emission of
optical phonons. If electrons are distributed near the top of the passive
region, a temporal negative absorption appears to be dominant in the
intersubband response. This is due to the effective broadening of the upper
level state under the optical phonon emission. We have then considered the
amplification of the ground mode in a THz waveguide with a multiquantum well
placed at the center of the cavity. A huge increase of the probe signal is
obtained, which permits the temporal stimulated emission regime of the
photoexcited electrons in the THz spectral region.
\end{abstract}

\maketitle

\section{Introduction}

The temporal emission of THz radiation due to the coherent oscillations of
electrons under ultrafast interband excitation have been studied during last
decades (see \cite{1} for review). The duration of the generated THz pulse
is in the order of picoseconds due to effective relaxation of the coherent
response. Furthermore, the steady-state spontaneous and stimulated emissions
from different semiconductor structures under an electric field pump have
been demonstrated (see \cite{2} and Refs. therein). However, the character
of the temporal evolution of photoexcited electrons during the non-coherent
stage of the relaxation is not completely understood yet. After the initial
photoexcitation and emission of the optical phonon cascade, which takes
place in a picosecond time interval, which have been considered in details
(see \cite{1,3,4,5} and Refs. therein), a non-equilibrium distribution
appears in the passive region, with the energy $\varepsilon_{ex}$ less than
the optical phonon energy, $\hbar \omega _{o}$ \cite{6}. A temporal
evolution of this distribution takes place during a nanosecond time
interval. This evolution is caused by the quasi-elastic scattering of
electrons with acoustic phonons. Due to the partial inversion of this
distribution, a set of peculiarities of the magnetotransport coefficients,
such as \textit{the total negative conductivity}, or the negative cyclotron
absorption, appears \cite{7}.

For the wide quantum well (QW) case under consideration, with two levels in
the passive region, the character of the evolution appears to be more
complicate due to the interlevel scattering. The regime of \textit{negative
absorption} is possible due to a more effective broadening of the absorption
(dashed arrow in Fig.1$a$) in comparison with the intersubband emission
(solid arrow). Such regime appears because the absorption process involves
the state in the active region for which the optical phonon emission is
allowed. Thus, a question arises about the temporal stimulated emission
coming from a wide multiple quantum well (MQW) structure placed at the
center of THz waveguide, as it is shown in Fig. 1$b$. In this paper we
consider the amplification of the probe ground mode in the THz resonator
caused by the temporal negative absorption described above. We have found a
huge amplification of the probe signal, so that it may be concluded that a
temporal stimulated emission takes place in the THz waveguide with weak
cavity losses and with the MQW structure placed at the center.

\begin{figure}[tbp]
\begin{center}
\includegraphics{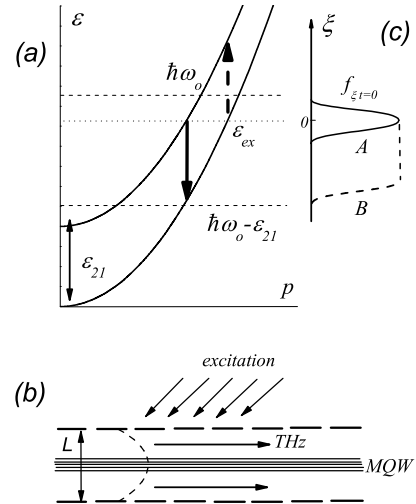}
\end{center}
\par
\addvspace{-1 cm}
\caption{Scheme of the intersubband transitions for electrons with energies
near the top of the passive region (the stimulated emission and absorption
are shown by the solid and dashed arrows, respectively) ($a$), the geometry
of the THz waveguide with the MQW structure in the center ($b$), and the
initial distributions over the passive region for the cases $A$ and $B$ ($c$%
).}
\label{fig.1}
\end{figure}

Our study is based on the general kinetic equation for the distribution
function in the conduction $c$-band, $f_{\alpha t}$, written in the
following form: 
\begin{equation}
\frac{\partial f_{\alpha t}}{\partial t}=G_{\alpha t}+J(f|\alpha t),
\end{equation}%
where $G_{\alpha t}$ and $J(f|\alpha t)$ are the photogeneration rate and
collision integral for the $c$-band state $\alpha $, respectively. The
generation rate in (1) is given by \cite{4}: 
\begin{equation}
G_{\alpha t}=w_{t}^{2}\sum_{l}g_{l}\delta _{\Delta _{l}}\left( \varepsilon
_{\alpha }-\varepsilon _{l}\right)
\end{equation}%
where $w_{t}$ is the form-factor of the excitation field with the pulse
duration $\tau _{p}$, the factor $g_{l}$ describes the relative
contributions of different valence $v$-band states (heavy and light holes
confined in the wide QW). The shape of the photoexcited distribution is
given by the Gaussian function $\delta _{\Delta }(E)=\exp [-(E/\Delta
)^{2}]/(\sqrt{\pi }\Delta )$ with the width $\Delta $ and centred at $E=0$.
The width of these function is determined by two different broadening
processes: the spreading due to the energy-time uncertainty relation and the
anisotropy of the valence band. These values are estimated by the energy
values $\hbar /\tau _{p}$ and $\beta \varepsilon _{l}$, respectively, where $%
\beta \ll 1$ is the anisotropy parameter. Under the optical phonon emission
an additional broadening appears. Such a broadening is caused by the weak
phonon dispersion, and characterized by the energy $\delta \varepsilon
_{opt} $.

Thus, the initial conditions in the passive region, after the picosecond
stage of the evolution due to optical phonon emission, are obtained from
Eqs. (1,2) in the following manner: 
\begin{equation}
f_{j\varepsilon t=0}=\sum_{l}\frac{n_{l}^{\scriptscriptstyle(j)}}{\rho _{%
\scriptscriptstyle2D}}\delta _{\Delta _{l}^{\scriptscriptstyle%
(j)}}(\varepsilon -\varepsilon _{l}^{\scriptscriptstyle(j)}).
\end{equation}%
Here $\rho _{\scriptscriptstyle2D}$ is the 2D density of states, the total
broadening energy is determined as $\Delta _{l}^{\scriptscriptstyle(j)}\sim 
\mathrm{max}(\hbar /\tau _{p},~\beta \varepsilon _{l},~\delta \varepsilon
_{opt})$, and $j=1,2$ are the corresponding two levels in the passive
region. Below, we will concentrate solely on two models of the initial
distribution: a narrow Gaussian peak (case $A$) and a flat distribution over
the interval $\left( \hbar \omega _{o}-\varepsilon _{21},\hbar \omega
_{o}\right) $ (case $B$), which are shifted from the boundary of the passive
region, as Fig. 1$c$ shows. Both distributions are relaxed between the
sub-bands and to the bottom of the passive region. The temporal intersubband
response is determined by the interlevel redistribution of the population.

The analysis we will carry out next is divided in two sections. The temporal
evolution of the distribution over the two subbands is described in Sec. II,
including the consideration of the resonant intersubband response. The
results for the transient amplification of a probe mode in the THz waveguide
are given in Sec. III, with our conclusions presented in Sec. IV. The
Appendix contains the microscopical evaluation of the broadening energy for
the intersubband transitions.

\section{Temporal evolution of the electronic distribution}

We shall now turn to consider the temporal evolution of the photoexcited
electrons in the passive region caused by the quasielastic scattering with
acoustic phonons. Since the above-discussed initial distributions are
isotropic over the 2D-plane, one has to consider the energy dependent
distribution functions $f_{1,2\varepsilon t}$ governed by the system of
kinetic equations: 
\begin{eqnarray}
\frac{\partial f_{1\varepsilon t}}{\partial t} &=&J(f|1\varepsilon t)-\nu
_{\varepsilon -\varepsilon _{21}}\left( f_{1\varepsilon t}-f_{2\varepsilon
-\varepsilon _{21}t}\right) ,  \nonumber \\
\frac{\partial f_{2\varepsilon t}}{\partial t} &=&J(f|2\varepsilon t)+\nu
_{\varepsilon +\varepsilon _{21}}\left( f_{1\varepsilon +\varepsilon
_{21}t}-f_{2\varepsilon t}\right) .
\end{eqnarray}%
Moreover, these equations are written for the intervals $\left( 0,
\hbar\omega_{o}\right)$ and $\left(\varepsilon_{21}-\hbar\omega _{o},
\hbar\omega_{o}\right)$ for the first and second subbands, respectively. The
interlevel relaxation frequency is introduced here as 
\begin{eqnarray}
\nu _{\varepsilon } &=&\theta \left( \varepsilon \right) \frac{2\pi }{\hbar }%
\int_{0}^{2\pi }\frac{d\varphi }{2\pi }\sum_{\mathbf{p}^{\prime }\mathbf{,}%
q_{\perp }}\left\vert C_{\scriptscriptstyle Q}\right\vert ^{2}  \nonumber \\
&&\times \left\vert \left\langle 2|e^{-iq_{\bot }z}|1\right\rangle
\right\vert ^{2}\left( 2N_{\scriptscriptstyle Q}+1\right) \delta \left(
\varepsilon -\varepsilon ^{\prime }\right) ,
\end{eqnarray}%
where $C_{Q}$ is the matrix element for the bulk electron-phonon interaction
with the acoustic phonons. The quasielastic collision integral in the $j$th
subband is reached as 
\begin{equation}
J(f|j\varepsilon t)=\frac{\partial }{\partial \varepsilon }\left(
D_{j\varepsilon }\frac{\partial f_{j\varepsilon t}}{\partial \varepsilon }%
+V_{j\varepsilon }f_{j\varepsilon t}\right) .
\end{equation}%
The diffusion and drift coefficients in (6), $D_{j\varepsilon }$ and $%
V_{j\varepsilon }$, are determined as follows: 
\begin{eqnarray}
\left\vert 
\begin{array}{l}
D_{j\varepsilon } \\ 
V_{j\varepsilon }%
\end{array}%
\right\vert &=&\frac{\pi \rho _{\scriptscriptstyle2D}}{\hbar }\int_{0}^{2\pi
}\frac{d\varphi }{2\pi }\int_{-\infty }^{\infty }\frac{dq_{\bot }}{2\pi }%
V\left\vert C_{\scriptscriptstyle Q}\right\vert ^{2} \\
&&\times \left\vert \left\langle j|e^{-iq_{\bot }z}|j\right\rangle
\right\vert ^{2}\left\vert 
\begin{array}{l}
\left( 2N_{\scriptscriptstyle Q}+1\right) (\hbar \omega _{\scriptscriptstyle %
Q})^{2}/ 2 \\ 
~~~~~~~~\hbar \omega _{\scriptscriptstyle Q}%
\end{array}%
\right\vert ,  \nonumber
\end{eqnarray}
where $N_{\scriptscriptstyle Q}=\left[ \exp \left( \hbar \omega
_{Q}/T\right) -1\right] ^{-1}$. A similar description of the quasi-elastic
relaxation in a two-level system was discussed in Ref. \onlinecite{8}.

Since the elastic interlevel relaxation is the dominant process for the
energy interval $(\varepsilon _{21},\hbar \omega _{o})$, one obtains $%
f_{1\varepsilon t}\approx f_{2\varepsilon -\varepsilon _{21}t}\simeq
f_{\varepsilon t}$ if $t$ exceeds the interlevel relaxation time. Using the
energy variable $\xi =\varepsilon -\hbar \omega _{o}$ we determine the
distribution function $f_{\xi +\hbar \omega _{o}t}\equiv f_{\xi t}$ from the
diffusion-drift equation for the two-level zone of the passive region, $%
0>\xi >\varepsilon _{21}-\hbar \omega _{o}$: 
\begin{equation}
\frac{\partial f_{\xi t}}{\partial t}=\frac{\partial }{\partial \xi }\left(
D_{\xi }\frac{\partial f_{\xi t}}{\partial \xi }+V_{\xi }f_{\xi t}\right) ,
\end{equation}%
where $D_{\xi }$ and $V_{\xi }$ are the energy diffusion and the drift
coefficients determined by Eq. (7). These coefficients are shown in Fig. 2
for a $GaAs$-based QW which is $320$ \AA\ wide, with an interlevel distance $%
\varepsilon _{21}=15$ meV and for two different temperatures $T=4.2$ K and $%
T=20$ K. Thus, the normalization coefficients $D_{\hbar \omega _{o}}\equiv
D_{\xi =0}=3.91\times 10^{9}$ meV$^{2}$/s and $9.55\times 10^{9}$ meV$^{2}$%
/s for $T=4.2$ K and $T=20$ K, respectively. $\ $Another coefficient, which
is independent on the temperature, is $V_{\hbar \omega _{o}}\equiv V_{\xi
=0}=5.22$ $\times 10^{9}$ meV/s. Note, that $D/V\simeq T$ for $T=20$ K while
a visible distinction of energy dependencies appears at $T=4$ K. We use
below the energy-independent coefficients $V=$ $3.5\times 10^{9}$ meV/s, and 
$D=2.2\times 10^{9}$ meV$^{2}$/s and $6.2\times 10^{9}$ meV$^{2}$/s for $%
T=4.2$ K and $T=20$ K, respectively. This approximation is valid within an
accuracy better than 20\% for the numerical parameters in the interval $%
\left(\hbar\omega_{o}-\varepsilon_{21},\hbar\omega_{o}\right)$ considered
below. Another important parameter is the interlevel relaxation time, $\nu
_{\varepsilon } ^{-1}$, which was assumed to be shorter than the time scales
of Figs. 3,4. Actually, for the above used parameters, $\nu _{\varepsilon
}^{-1}$ varies from $0.21$ ns to $0.19$ ns over the passive region. 
\begin{figure}[tbp]
\begin{center}
\includegraphics{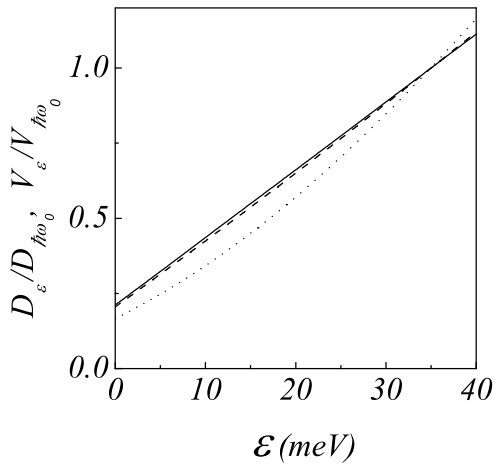}
\end{center}
\par
\addvspace{-1 cm}
\caption{Normalized diffusion and drift coefficients versus energy. Solid
and dotted curves: diffusion coefficient for $T=$4.2 K and $T=$20 K
respectively. Dashed line: drift coefficient which is independent of $T$.}
\label{fig.2}
\end{figure}

Equation (8) may be considered with the zero boundary conditions at $\xi
\rightarrow \pm \infty $ if $\hbar \omega _{o}-\varepsilon _{21}>\varepsilon
_{21}$. The initial distribution is located in the region $\xi <0$, and
electrons do not reach the bottom of the second level, i.e., $\xi
>-\varepsilon _{21}$. For the case $A$ we use the initial condition: $f_{\xi
t=0}=n_{ex}\delta _{\Delta }(\xi -\xi _{ex})/2\rho _{\ss 2D}$, where $n_{ex}$
is the total excited concentration, $\xi _{ex}\simeq -9$ meV is the
excitation energy, and the half-width of the peak is equal to 2.5 meV. Thus,
the solution for Eq.(8) may be written as the moving peak 
\begin{equation}
f_{\xi t}=\frac{n_{ex}}{2\rho _{\ss 2D}}\delta _{\Delta _{t}}(\xi +Vt-\xi
_{ex})
\end{equation}%
with the time-dependent half-width $\Delta _{t}=\sqrt{\Delta ^{2}+4Dt}$. For
the case $B$, the broadening of the stepped distribution is equal to 2.5
meV. The solution is given by the integral of the initial distribution
multiplied by a similar to (2) form-factor. This solution is plotted in
Fig.3 for different times. 
\begin{figure}[tbp]
\begin{center}
\includegraphics{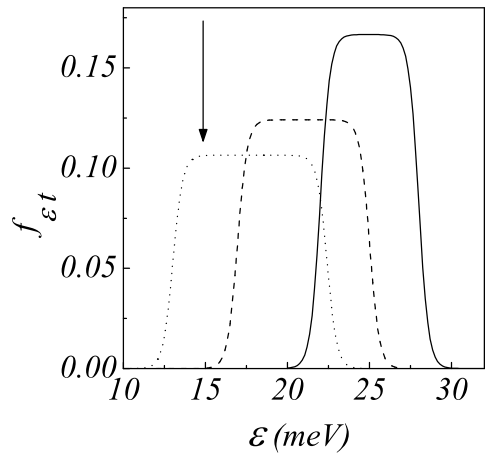}
\end{center}
\par
\addvspace{-1 cm}
\caption{Evolution of the electron distribution for the case $B$. Solid,
dashed, and dotted curves correspond to $t=$0 ns, 1.6 ns, and 3.2 ns
respectively. Vertical arrow indicates the position of $\protect\varepsilon %
_{21}$.}
\label{fig.3}
\end{figure}

In order to describe the temporal negative absorption, we need the
concentration over the negative absorption region $\left( \hbar \omega
_{o}-\varepsilon _{21},\hbar \omega _{o}\right) $ defined as $n_{t}=2\rho _{%
\ss 2D}\int_{-\varepsilon _{21}}^{0}d\xi f_{\xi t}$. The temporal evolution
of $n_{t}$, obtained with the solutions of Eq.(8) for the cases $A$ and $B$,
and the QW parameters detailed above, is shown in Fig.4. 
\begin{figure}[tbp]
\begin{center}
\includegraphics{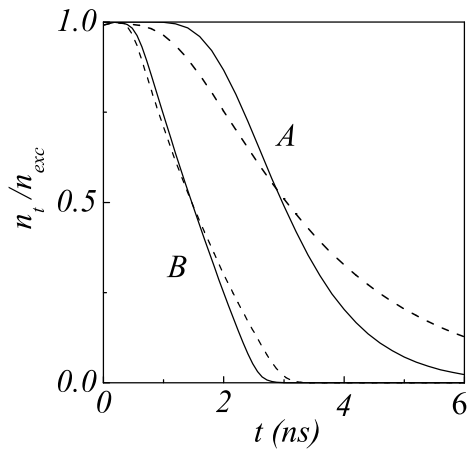}
\end{center}
\par
\addvspace{-1 cm}
\caption{Temporal evolution of the concentration for the distributions $A$
and $B$, at $T=$4.2 K and 20 K (solid and dashed curves, respectively).}
\label{fig.4}
\end{figure}

The temporal evolution of the resonant absorption of the non-equilibrium
electrons is described by the real part of the conductivity (see \cite{8}
and Appendix): 
\begin{equation}
Re\sigma _{\delta \varepsilon ,t}=\frac{e^{2}|v_{\ss \perp }|^{2}}{%
\varepsilon _{21}/\hbar }\frac{n_{t}}{2}\left( \frac{\Gamma }{\delta
\varepsilon ^{2}+\Gamma ^{2}}-\frac{\gamma }{\delta \varepsilon ^{2}+\gamma
^{2}}\right) .
\end{equation}%
Here $\delta \varepsilon =\hbar \omega -\varepsilon _{21}$ is the detuning
energy, $v_{\ss \perp }$ is the intersubband velocity matrix element, $%
\Gamma $ and $\gamma $ are the broadening energy values due to the optical
phonon emission and the elastic scattering, which correspond to the
absorption and stimulated emission processes. Using Eq.(A11) and the
parameters given above we make an estimate of $\Gamma \sim $ $1.3(1.5)$ meV
[for $\xi _{ex}=-4(-9)$ meV, respectively], so that $\Gamma \gg \gamma $ and
one obtains $Re\sigma _{\delta \varepsilon =0,t}<0$. The contribution to $%
Re\sigma _{\delta \varepsilon ,t}$ from electrons distributed over the
region $-\varepsilon _{21}>\xi >\varepsilon _{21}-\hbar \omega _{o}$ is
equal to zero due to the same broadening of the absorption and emission
processes. We have also supposed that electrons are absent from the region $%
\xi <\varepsilon _{21}-\hbar \omega _{o}$, where only the absorption due to
transitions $1\rightarrow 2$ is possible.

\section{Transient amplification}

We then consider the amplification of a probe THz mode in the ideal
waveguide of width $L$ due to the temporal negative absorption discussed
above. The transverse electric field in the resonator $E_{zt}^{\ss \perp
}\exp (-i\omega t+ikx)$ is ruled by the wave equation: 
\begin{equation}
\left( \frac{\partial ^{2}}{\partial z^{2}}-\kappa ^{2}\right) E_{zt}^{\ss %
\perp }+i\epsilon \frac{2\omega }{c^{2}}\frac{\partial E_{zt}^{\ss \perp }}{%
\partial t}=0
\end{equation}%
with $\kappa ^{2}=k^{2}-\epsilon \omega ^{2}/c^{2}$, where $\epsilon $ is
the dielectric permittivity supposed to be uniform across the structure. The
boundary conditions at $z=\pm L/2$ takes the form $E_{z=\pm L/2,t} ^{\ss %
\perp }=0$. At the center of the resonator ($z=0$), where the MQW is placed,
one has to use: 
\begin{equation}
\left. E_{zt}^{\ss \perp }\right\vert _{-0}^{0}=0,~~~~\left. \frac{\partial
E_{zt}^{\ss \perp }}{\partial z}\right\vert _{-0}^{0}\simeq -iN\frac{4\pi
\omega }{c^{2}}\sigma _{\delta \varepsilon ,t}E_{z=0t}^{\ss \perp },
\end{equation}%
where $N$ is the number of wells in the structure. The initial condition for
the ground mode propagating along the resonator is: $E_{zt=0}^{\ss \perp
}=E\cos (\pi z/L)$.

Taking into account the slowness of the temporal evolution under the
condition $\alpha =N\pi e^{2}|v_{\ss \bot }|^{2}Ln_{ex}/\gamma c^{2}\ll 1$
and restricting ourselves to the resonant case, $\delta \varepsilon =0$, we
obtain the solution of Eqs. (11, 12) as $E_{zt}^{\ss \perp }=\mathcal{E}%
_{t}\cos (\pi z/L)$ where the time-dependent field is governed by: 
\begin{equation}
\frac{\partial \mathcal{E}_{t}}{\partial t}=\frac{n_{t}}{\tau ^{\ast }n_{ex}}%
\mathcal{E}_{t},~~~~\frac{1}{\tau ^{\ast }}=\frac{4\alpha c^{2}}{\epsilon
\omega _{21}L^{2}},
\end{equation}%
with the initial condition: $\mathcal{E}_{t=0}=E$. 
\begin{figure}[tbp]
\begin{center}
\includegraphics{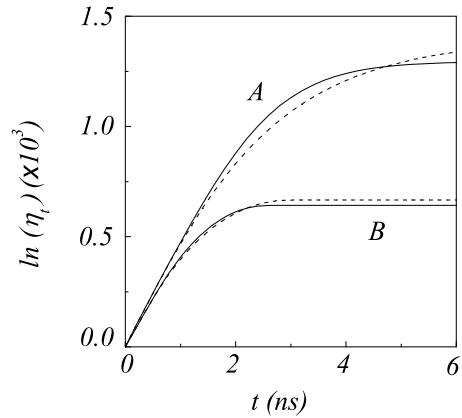}
\end{center}
\par
\addvspace{-1 cm}
\caption{Transient amplification $\protect\eta _{t}$ for the same conditions
as in Fig.4.}
\label{fig.5}
\end{figure}

After averaging over the resonator, the Poynting vector along OX-direction
is introduced as 
\begin{equation}
S_{t}=\frac{kc^{2}}{2\pi \omega }\int dz|\mathcal{E}_{zt}|^{2}
\end{equation}%
and the temporal amplification is determined as the ratio of (14) to the
Poynting vector at $t=0$: 
\begin{equation}
\eta _{t}=\frac{S_{t}}{S_{t=0}}=\exp \left( 2\int_{0}^{t}\frac{dt^{\prime }}{%
\tau ^{\ast }}\frac{n_{t^{\prime }}}{n_{ex}}\right) .
\end{equation}
In Fig. 5 we plotted the logarithm of the amplification coefficient versus
time for a five-layers MQW with $\gamma =0.2$ meV under the photoexcited
concentration $n_{ex}=10^{10}$ cm$^{-2}$. We have also used $\epsilon =12.9$
and $L=19$ $\mu $m, corresponding to the ground mode propagation along the
THz waveguide. One can see the saturation of $\ln \eta _{t}$ when $%
n_{t}/n_{ex}$ approaches to zero (cf. Fig.4). The maximal value of $\ln \eta
_{t}$ is around\ $10^{3}$, i.e., a huge temporal amplification of the probe
signal has been obtained. Thus, the stimulated regime of the emission should
be realized in the THz waveguide with a weak damping (such waveguides have
been studied recently, see Ref. \onlinecite{9} and references therein).

\section{Conclusions}

In summary, we have considered a new non-coherent transient mechanism of the
stimulated emission in the THz spectral region. The mechanism neither
requires coherent response nor inverted distribution but it rather appears
due to the different broadening of emission and absorption contributions in
Eq. (10).

The consideration performed here is based on several assumptions. Rather
than using a microscopic calculation of the photogeneration, we have used
two models for initial conditions ($A$ and $B$) to obtain the temporal
evolution of the distribution. We have also neglected the energy
dependencies of $D_{\xi }$ and $V_{\xi }$ in Eq. (8). In Eq. (10), which
describes the intersubband response, we have neglected the Coulomb
renormalization effect and taken into account that $\Gamma \gg \gamma $ as
it is demonstrated in the Appendix. We also neglected nonparabolicity of the
electron dispersion law and the effect of the spatial confinement on phonons
and electron-phonon interaction \cite{10}. These approximations are
generally accepted for the $GaAs/AlGaAs$ heterostructures with the
parameters used here. In addition, instead of a stimulated emission effect
due to the transient negative absorption, we have calculated the
amplification of a probe signal, and the role of the waveguide losses is not
considered here.

To conclude, we have found a huge amplification due to the temporal negative
absorption under intersubband transitions of photoexcited electrons in a
wide MQW structure placed at the center of a THz resonator. We expect that
the presented analysis motivates an experimental treatment of the transient
stimulated emission both for the system considered and for another
heterostructures with closely-spaced levels (stepped QW, tunnel-coupled
structures, etc.).

\begin{acknowledgments}
This work has been supported in part by the Consejer\'{\i}a de Educaci\'{o}%
n, Cultura y Deportes. Gobierno Aut\'{o}nomo de Canarias.
\end{acknowledgments}

\appendix

\section{Broadening of intersubband transitions}

Below we present the evaluation for the resonant intersubband conductivity
given by Eq. (10). We consider the interaction of electrons with optical
phonons at zero temperature and demonstrate that only the broadening of the 1%
$\rightarrow $2 transition, which is responsible for the absorption, takes
place in the passive region while the 2$\rightarrow $1 transition remains
narrow (the broadening is determined by the weak elastic scattering).

Within the framework of nonequilibrium diagram technique \cite{11,3}, one
describes the transient response on the resonant probe field $E_{\ss \bot
}\exp (-i\omega t)$ based on the linear combination of Green's functions: 
\begin{equation}
\hat{F}_{t_{1}t_{2}}+\hat{G}_{t_{1}t_{2}}^{\ss A}-\hat{G}_{t_{1}t_{2}}^{\ss %
R}\simeq \int \frac{d\varepsilon }{2\pi \hbar }e^{\frac{i}{\hbar }%
\varepsilon (t_{1}-t_{2})}\hat{\mathcal{F}}_{\varepsilon \frac{t_{1}+t_{2}}{2%
}}+\widehat{\delta \mathcal{F}}_{t_{1}t_{2}}.
\end{equation}%
Here $\hat{\mathcal{F}}_{\varepsilon t}=2f_{\varepsilon t}(\hat{G}%
_{\varepsilon }^{\ss A}-\hat{G}_{\varepsilon }^{\ss R})$ describes the
temporal evolution of energy distribution $f_{\varepsilon t}$ governed by
the quasiclassic kinetic equation (1), $\hat{G}^{\ss A,R}$ is the advanced ($%
A$) or retarded ($R$) equilibrium Green's function, and $\widehat{\delta 
\mathcal{F}}_{t_{1}t_{2}}$ describes the linear response on the perturbation 
$\widehat{\delta h}\exp (-i\omega t)$. Using the $\varepsilon ,t$%
-representation we obtain the linearized equation for $\widehat{\delta 
\mathcal{F}}_{\varepsilon }\exp (-i\omega t)$ in the following form: 
\begin{eqnarray}
&&\hbar \omega \widehat{\delta \mathcal{F}}_{\varepsilon }-[\hat{h},\widehat{%
\delta \mathcal{F}}_{\varepsilon }]-\left( \widehat{\delta h}\hat{\mathcal{F}%
}_{\varepsilon -\hbar \omega /2,t}-\hat{\mathcal{F}}_{\varepsilon +\hbar
\omega /2,t}\widehat{\delta h}\right) ~~~  \nonumber \\
&=&\widehat{\delta \Omega }_{\varepsilon }\hat{G}_{\varepsilon -\hbar \omega
/2}^{\ss A}-\hat{G}_{\varepsilon +\hbar \omega /2}^{\ss R}\widehat{\delta
\Omega }_{\varepsilon }  \nonumber \\
&&+\hat{\Sigma}_{\varepsilon +\hbar \omega /2}^{\ss R}\widehat{\delta 
\mathcal{F}}_{\varepsilon }-\widehat{\delta \mathcal{F}}_{\varepsilon }\hat{%
\Sigma}_{\varepsilon +\hbar \omega /2}^{\ss A},
\end{eqnarray}%
where $\hat{\Sigma}_{\varepsilon }^{\ss R,A}$ is the $R$- or $A$-
self-energies and the integral contribution to this equation is given by 
\begin{equation}
\widehat{\delta \Omega }_{\varepsilon }\simeq \sum_{\mathbf{Q}}|C_{\ss Q}^{%
\ss (po)}|^{2}e^{i\mathbf{Q\cdot r}}\widehat{\delta \mathcal{F}}%
_{\varepsilon -\hbar \omega _{o}}e^{-i\mathbf{Q\cdot r}}.
\end{equation}%
Here $C_{\ss Q}^{\ss (po)}$ is the matrix element of the Fr\"{o}hlich
interaction with the bulk phonon mode characterized by the wave vector $%
\mathbf{Q}=(\mathbf{q}, q_{\ss \bot})$.

For the case of the resonant intersubband excitation, $|\hbar \omega
-\varepsilon _{21}|\ll \varepsilon _{21}$, only the component $(\widehat{%
\delta \mathcal{F}}_{\varepsilon })_{21,\mathbf{p}}\equiv \delta
F_{\varepsilon \mathbf{p}}$ appears to be essential. Using the non-diagonal
matrix element $\delta h_{21}=(ie/\omega )E_{\ss \bot }v_{\ss \bot }$ we
transform Eqs. (A2) and (A3) into 
\begin{eqnarray}
&&\left( \hbar \omega -\varepsilon _{21}+i\gamma _{p}\right) \delta
F_{\varepsilon \mathbf{p}}-\sum_{\mathbf{p}}w_{\mathbf{pp}^{\prime }}\delta
F_{\varepsilon -\hbar \omega _{o}\mathbf{p}^{\prime }}  \nonumber \\
&=&2\frac{ie}{\omega }E_{\ss \bot }v_{\ss \bot }\left[ f_{\varepsilon -\hbar
\omega /2}\left( G_{\varepsilon -\hbar \omega /2}^{\ss A}-G_{\varepsilon
-\hbar \omega /2}^{\ss R}\right) _{11,\mathbf{p}}\right.  \nonumber \\
&&\left. -f_{\varepsilon +\hbar \omega /2}\left( G_{\varepsilon +\hbar
\omega /2}^{\ss A}-G_{\varepsilon +\hbar \omega /2}^{\ss R}\right) _{22,%
\mathbf{p}}\right] ,
\end{eqnarray}%
where the broadening energy is given by 
\begin{eqnarray}
\gamma _{p} &=&i\left[ \left( \hat{\Sigma}_{\varepsilon +\hbar \omega /2}^{%
\ss R}\right) _{22,\mathbf{p}}-\left( \hat{\Sigma}_{\varepsilon +\hbar
\omega /2}^{\ss A}\right) _{11,\mathbf{p}}\right]  \nonumber \\
&\simeq &i\sum_{j\mathbf{Qp^{\prime }}}|C_{\ss Q}^{\ss (po)}|^{2}\left[ 
\frac{|\langle 2\mathbf{p}|e^{i\mathbf{Q\cdot r}}|j\mathbf{p^{\prime }}%
\rangle |^{2}}{\varepsilon +\hbar \omega /2-\hbar \omega _{o}+i\lambda }%
\right.  \nonumber \\
&&\left. -\frac{|\langle 1\mathbf{p}|e^{i\mathbf{Q\cdot r}}|j\mathbf{%
p^{\prime }}\rangle |^{2}}{\varepsilon -\hbar \omega /2-\hbar \omega
_{o}+i\lambda }\right]
\end{eqnarray}%
and $w_{\mathbf{pp^{\prime }}}$ is determined by a similar expression which
is not essential below. The lower Eq. (A5) is written in the Born
approximation and, from the formal point of view, $\lambda \rightarrow +0$.
If we take into account the elastic scattering, $\lambda $ should be
replaced by the elastic broadening energy.

The Fourier component of current density due to the intersubband transitions
under consideration is expressed through $\widehat{\delta F}_{\varepsilon 
\mathbf{p}}$ according to $\mathbf{j}_{\omega }=-iev_{\ss \bot }\int d 
\mathbf{p}\int d\varepsilon \widehat{\delta F}_{\varepsilon \mathbf{p}}
/(2\pi\hbar )^{2}$. Here we have replaced the perturbation of the density
matrix $\widehat{\delta \rho }_{t}$ using the relation: $\widehat{\delta
\rho }_{t}=(-i\hbar /2)\lim_{t_{1,2}}\widehat{\delta \mathcal{F}}%
_{t_{1}t_{2}}$. Performing the substitution $\delta F_{\varepsilon \mathbf{p}%
}=i2eE_{\ss \bot }v_{\ss \bot }\varphi _{\varepsilon \mathbf{p}}/\omega $ we
obtain the conductivity : 
\begin{equation}
\sigma _{\omega }=\frac{e^{2}|v_{\ss \bot }|^{2}}{\pi}\int\frac{ d \mathbf{p}%
}{(2\pi\hbar )^2}\int d\varepsilon \varphi _{\varepsilon \mathbf{p}}.
\end{equation}%
The function $\varphi _{\varepsilon \mathbf{p}}$ is governed by the integral
equation 
\begin{eqnarray}
&&\left( \hbar \omega -\varepsilon _{21}+i\gamma _{p}\right) \varphi
_{\varepsilon \mathbf{p}}-\sum_{\mathbf{p}}w_{\mathbf{pp}^{\prime }}\varphi
_{\varepsilon -\hbar \omega _{o}\mathbf{p}}  \nonumber \\
&=&f_{\varepsilon -\hbar \omega /2}\left( G_{\varepsilon -\hbar \omega /2}^{%
\ss A}-G_{\varepsilon -\hbar \omega /2}^{\ss R}\right) _{11,\mathbf{p}} 
\nonumber \\
&&-f_{\varepsilon +\hbar \omega /2}\left( G_{\varepsilon +\hbar \omega
/2}^{A}-G_{\varepsilon +\hbar \omega /2}^{R}\right) _{22,\mathbf{p}},
\end{eqnarray}%
which is obtained from Eq. (A4).

Next we write the resonant conductivity (A6) for the case of the $\delta $%
-like electron distribution $f_{\varepsilon }=(n_{ex}/2\rho _{\ss 2D})\delta
(\varepsilon _{ex}-\varepsilon )$ with the excitation energy $\varepsilon
_{ex}$ localized in the interval $(\hbar \omega _{o}-\varepsilon _{ex},\hbar
\omega _{o})$. Due to the weakness of the integral term in Eq.(A7) we obtain
the conductivity in the form: 
\begin{eqnarray}
\sigma _{\omega } &=&\frac{e^{2}|v_{\ss \bot }|^{2}}{2\pi }n_{ex}\int
d\varepsilon _{p}\left[ \frac{\left( G_{\varepsilon -\hbar \omega /2}^{\ss %
A}-G_{\varepsilon -\hbar \omega /2}^{\ss R}\right) _{11,\mathbf{p}}}{\hbar
\omega -\varepsilon _{21}+i\gamma _{p~|\varepsilon \rightarrow \varepsilon
_{ex}+\hbar \omega /2}}\right.  \nonumber \\
&&\left. -\frac{\left( G_{\varepsilon -\hbar \omega /2}^{\ss %
A}-G_{\varepsilon -\hbar \omega /2}^{\ss R}\right) _{22,\mathbf{p}}}{\hbar
\omega -\varepsilon _{21}+i\gamma _{p~|\varepsilon \rightarrow \varepsilon
_{ex}-\hbar \omega /2}}\right] .~~~~
\end{eqnarray}%
The spectral densities in the numerators, 
\begin{equation}
\left( G_{\varepsilon -\hbar \omega /2}^{A}-G_{\varepsilon -\hbar \omega
/2}^{R}\right) _{jj,\mathbf{p}}\simeq \frac{i2\lambda }{(\varepsilon
_{ex}-\varepsilon _{jp})^{2}+\lambda ^{2}},
\end{equation}%
are written through the $\delta $-functions if $\lambda \rightarrow +0$. As
a result, Eq.(A8) is transformed into Eq.(10) with the elastic broadening
energy in the second contribution while the optical phonon induced
broadening appears in the first term. The corresponding broadening energy is
determined as $\Gamma =Re\gamma _{p}|_{\varepsilon _{p}\rightarrow
\varepsilon _{ex},\varepsilon \rightarrow \varepsilon _{ex}+\hbar \omega /2}$%
, or 
\begin{eqnarray}
\Gamma &\simeq &\frac{2\pi ^{2}e^{2}\hbar \omega _{o}}{\epsilon ^{\ast }V}%
\sum_{q_{\bot }\mathbf{p}^{\prime }}\frac{|\langle 2|e^{iq_{\bot
}z}|1\rangle |^{2}}{|\mathbf{p}-\mathbf{p}^{\prime }|/\hbar ^{2}+q_{\bot
}^{2}}  \nonumber \\
&&\times \delta \left( \varepsilon _{ex}+\varepsilon _{21}-\hbar \omega
_{o}-\varepsilon _{1p^{\prime }}\right) _{|\varepsilon _{p}\rightarrow
\varepsilon _{ex}},
\end{eqnarray}%
where $V$ is the normalization volume and the effective dielectric
permittivity, $\epsilon ^{\ast }=(\epsilon_{o}-\epsilon_{\infty })/ \epsilon
_{o}\epsilon _{\infty }$, is introduced through the static and
high-frequency dielectric permittivities, $\epsilon_{o}$ and $%
\epsilon_{\infty }$. Performing the integration over the energy one obtains: 
\begin{equation}
\Gamma \simeq \frac{e^{2}\hbar \omega _{o}}{4\epsilon ^{\ast }}\rho _{\ss %
2D}\int dq_{\bot }\int_{0}^{2\pi }d\varphi \frac{|\langle 2|e^{iq_{\bot
}z}|1\rangle |^{2}}{q_{\bot }^{2}+Q_{\varphi }^{2}},
\end{equation}%
where the denominator contains the factor $Q_{\varphi }^{2}=(p_{ex}^{2}+%
\tilde{p}_{ex}^{2}-2p_{ex}\tilde{p}_{ex}\cos \varphi )/\hbar ^{2}$ with the
characteristic momenta $p_{ex}=\sqrt{2m\varepsilon _{ex}}$ and $\tilde{p}%
_{ex}=\sqrt{2m[\varepsilon _{ex}-(\hbar \omega _{o}-\varepsilon _{21})]}$.

\end{document}